\documentstyle[aps,pre,epsf,floats]{revtex}

\def\xvec{\text{{\bf x}}}
\def\vacancy{\text{\O}}

\begin{document}
\parskip 1mm
\twocolumn[\hsize\textwidth\columnwidth\hsize\csname@twocolumnfalse%
\endcsname
\title  {On possible experimental realizations of directed percolation}

\author {Haye Hinrichsen\\[2mm]}

\address{Theoretische Physik, Fachbereich 10, 
	 Gerhard-Mercator-Universit\"at Duisburg, 
	 D-47048 Duisburg, Germany\\
 }
\date   {October 28, 1999}
\maketitle

\begin{abstract}
Directed percolation is one of the most prominent
universality classes of nonequilibrium phase transitions
and can be found in a large variety of models.
Despite its theoretical success, no experiment is
known which clearly reproduces the critical exponents of 
directed percolation. The present work compares suggested
experiments and discusses possible reasons why
the observation of the critical exponents of directed
percolation is obscured or even impossible.
\end{abstract}

]
%
%
%
%
\tableofcontents

\section{Introduction}

Physical phenomena far from thermal equilibrium are very
common in nature. For example, many systems are subjected to an
external flow of energy or particles which keeps them away from
equilibrium. Similarly, dynamic systems starting with a nonequilibrium
initial state may need a long time to reach thermal equilibrium.
Theories of systems out of equilibrium are more difficult than
equilibrium statistical mechanics since the 
partition sum is no longer given by the Gibbs ensemble.
On the other hand, nonequilibrium systems may exhibit a
potentially richer behavior than systems at thermal equilibrium.
Therefore, the study of nonequilibrium phenomena is a field
of growing interest, both theoretically and experimentally.

A particularly interesting topic is the investigation
of phase transitions far from equilibrium~\cite{MarroDickman99}. 
The theoretical interest in nonequilibrium 
phase transitions mainly stems from
the emergence of {\it universal} features of the
associated critical behavior. 
The concept of universality was originally introduced
by experimentalists in the context of equilibrium systems in order to
describe the observation that order parameters
of various apparently unrelated systems may display 
the same type of singular behavior near the transition.
These singularities are associated with a certain set of
critical exponents which characterizes the universality class 
of the transition. The subsequent development of powerful
theoretical concepts such as scaling, renormalization group,
and conformal invariance supported this hypothesis 
and established universality as a paradigm 
of equilibrium statistical mechanics.

Because of this success, theoretical physicists are nowadays
trying to transfer the idea of universality to nonequilibrium phase
transitions. However, in the nonequilibrium case the 
emerging picture remains less clear. Although 
universality certainly exists on the level of simple models,
the experimental evidence of universal behavior under nonequilibrium
conditions is still very poor. Therefore, it is not yet known to what 
extent the concept of universality can be applied to 
nonequilibrium critical phenomena.

An important example is the universality
class of {\em Directed Percolation} (DP)
which describes continuous phase
transitions from a spreading (wet)
phase into an absorbing (dry) state~\cite{Kinzel83}.
The DP universality class is extremely robust
with respect to the choice of the dynamic rules
and covers a large variety of
models with applications ranging from catalytic
reactions~\cite{ZGB86} and interface growth~\cite{TangLeschhorn92}
to turbulence~\cite{Pomeau86}. The
observed robustness led Janssen and 
Grassberger~\cite{DPConjecture} to the conjecture
that a continuous phase transition
from a fluctuating active phase into a single
absorbing state should belong to the DP universality
class, provided that the model uses short-range
dynamics without special attributes such as additional
symmetries or quenched randomness.
In fact, DP is the canonical universality class
for nonequilibrium phase transitions into absorbing states.
Thus it may be as important 
as the Ising universality class in 
equilibrium statistical mechanics.

Despite this success in theoretical statistical physics, 
the critical behavior of DP, 
especially the set of critical exponents, has
not yet been confirmed experimentally.
The lack of experimental evidence is indeed surprising,
especially since a large number of possible experimental
realizations have been suggested in the past.
As Grassberger emphasizes in a summary 
on open problems in DP~\cite{Grassberger96}:
\begin{quote}
{\it "...there is still no experiment where the critical 
behavior of DP was seen. This is a very strange situation
in view of the vast and successive theoretical efforts made 
to understand it. Designing and performing
such an experiment has thus top priority in my 
list of open problems."}.
\end{quote}

\noindent
The aim of the present work is to review the most
important experiments which have been suggested so far 
and to discuss their specific problems which 
could obscure the verification of the critical exponents. 
After a brief introduction to DP in
Sect.~\ref{DPSection} we will first discuss certain catalytic
reactions on two-dimensional surfaces which mimic the
dynamic rules of DP. As described in Sect.~\ref{GrowthSection},
DP may also be realized in certain wetting experiments where
the interface between air and liquid undergoes a depinning
transition. Another possible realization may be 
provided by systems of flowing sand on 
an inclined plane (see Sect.~\ref{SandSection}). 
However, in none of these experiments
the predicted critical exponents could be confirmed
convincingly.

What might be the reason for the apparent lack of experimental
evidence? It seems that the basic features of DP, which can
easily be implemented on a computer, are quite difficult to 
realize in nature. One of these theoretical 
assumptions is the existence of an absorbing state. 
In real systems, however, a perfect non-fluctuating state cannot 
be realized. For example, a poisoned catalytic
surface is not completely frozen, it will rather always
be affected by small fluctuations.
Although these fluctuations are strongly suppressed, they
could still be strong enough to `soften' the transition,
making it impossible to quantify the critical exponents.

Another reason might be the influence of 
{\it quenched disorder} due to spatial or temporal 
inhomogeneities. In most experiments frozen randomness
is expected to play a significant role. For example, 
a real catalytic surface is  not fully homogeneous 
but characterized by certain defects leading to 
spatially quenched disorder. As shown in Ref.~\cite{WACH98}, 
this type of disorder may affect or even destroy the 
critical behavior of DP. Technical details concerning
quenched disorder are summarized in the Appendix.

\section{Directed Percolation}
\label{DPSection}

%
%
\begin{figure}
\epsfxsize=85mm
\centerline{\epsffile{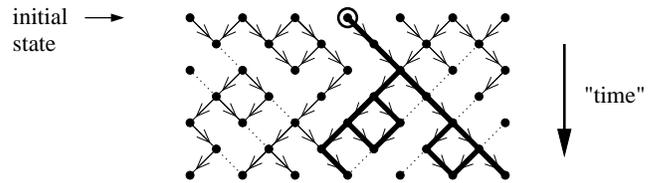}}
\vspace{2mm}
\caption{\small
\label{FIGDP}
Directed bond percolation on a diagonal square lattice.
Open (closed) bonds are represented as solid (dashed)
lines. The spreading agent, introduced at the site marked
by the circle, flows downwards through open bonds, 
generating a certain cluster (bold bonds).
}
\end{figure}

Directed percolation was introduced by
Broadbent and Hammersley~\cite{BroadbentHammersly57} 
in 1957 as an anisotropic variant of 
ordinary percolation~\cite{StaufferAharony92}.
The aim was to describe the connectivity of directed
random media such as porous rocks in a gravitational
field. Assuming that water can only move downwards,
the permeability of a rock will depend on the average
connectivity of its pores. One of the simplest models is
{\it directed bond percolation}. As shown in Fig.~\ref{FIGDP}, 
neighboring sites (pores) of a diagonal square lattice 
are connected by bonds (channels) which are open with 
probability $p$ and otherwise closed. Because of the
gravitational field, the bonds function as `valves'
wherefore the spreading agent can only percolate
along the given direction, as indicated by the arrows.

\begin{figure}
\epsfxsize=85mm
\centerline{\epsffile{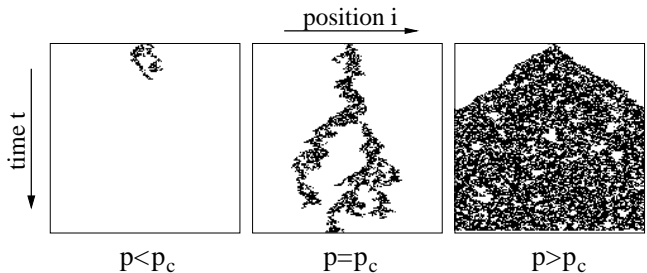}}
\vspace{2mm}
\caption{\small
\label{FIGDPDEMO}
Typical DP cluster starting from a single seed
below, at, and above criticality.
}
\end{figure}

Depending on the actual configuration of open bonds, each site
generates a certain {\it cluster} of connected sites.
A cluster of this kind would correspond to the maximal spreading
range if the water was injected into a single pore.
Below a certain threshold $p<p_c$ all clusters are finite,
i.e., the material is impermeable on large scales. 
However, above the critical value a 
cluster may become infinite so that water can percolate over
arbitrarily long distances (cf. Fig~\ref{FIGDPDEMO}).

Regarding the given direction as `time', DP may be 
interpreted as a $d$+1-dimensional dynamic process 
describing the spreading of some non-conserved agent. 
For example, we may enumerate the sites in Fig.~\ref{FIGDP}
horizontally by a spatial coordinate $i$ and vertically 
by a discrete time variable $t$. Then, for a given
state at time $t$, we can determine the state at time
$t+1$ by means of certain stochastic updates~\cite{DK}.

Interpreting active sites as particles $A$ and inactive sites
as vacancies $\vacancy$, the particles can either destroy
themselves or produce an offspring. Moreover, if two 
particles reach the same site, they coagulate to a 
single particle. Therefore, DP may be regarded as a 
reaction-diffusion process
\begin{equation}
\label{DPReactionDiffusion}
\begin{array}{ll}
\text{diffusion:} & \vacancy+A \rightarrow A+\vacancy \ , \\
\text{self-destruction:} & A \rightarrow \vacancy \ , \\
\text{offspring production:} & A \rightarrow A+A \ , \\
\text{coagulation:} & A+A \rightarrow A \ . \\
\end{array}
\end{equation}
Depending on the ratio between offspring production and
self destruction, the process may either remain active
or reach the empty state from where it cannot escape.
This is the so-called absorbing state of DP systems.
For a special realization of DP, the so-called contact
process~\cite{Harris74}, the existence of a continuous transition between 
survival and extinction could be proven rigorously~\cite{Liggett85}.
Near the transition the stationary density of active sites 
vanishes as a power law
\begin{equation}
\rho_{stat} \sim (p-p_c)^\beta \,,
\end{equation}
where $\beta$ is the critical exponent associated with the
order parameter $\rho$. Moreover, the DP process is 
characterized by a spatial and a temporal correlation length
diverging at the transition as
\begin{equation}
\xi_\perp     \sim |p-p_c|^{-\nu_\perp} \,, \qquad
\xi_\parallel \sim |p-p_c|^{-\nu_\parallel} \,.
\end{equation}
As already mentioned, a large variety of models
show essentially the same properties at the transition,
forming the DP universality class. The DP class corresponds
to a specific field theory~\cite{FT} and is characterized by 
the three critical exponents $\beta,\nu_\perp,\nu_\parallel$.
Despite its simplicity, DP has not yet been solved exactly.
However, the critical exponents can easily be estimated by
computer simulations. The most accurate estimates are 
summarized in Table I. Notice that in $d > 4$ spatial
dimensions fluctuations become irrelevant so that the 
exponents are given by their mean field values.
%
%
%
%

\section{Catalytic reactions}
\label{CatalyticSection}

\begin{figure}
\epsfxsize=85mm
\centerline{\epsffile{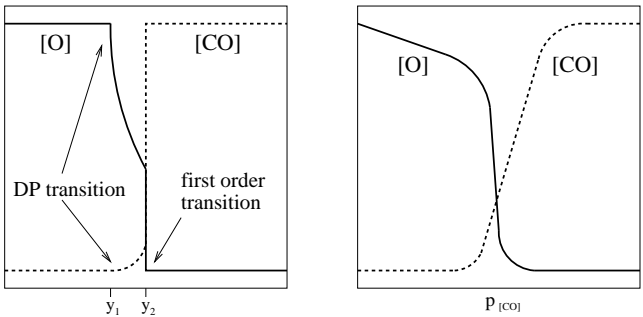}}
\vspace{1mm}
\caption{\small
\label{FIGZGB}
Catalytic reactions in theory (left) and experiment (right).
The schematic graphs show the concentrations of oxygen (solid line) and
carbonmonoxide (dashed line) as a function of the CO adsorption rate.
}
\end{figure}

It is well known that under specific conditions certain
catalytic reactions mimic the microscopic rules of 
DP models. For example, the Ziff-Gulari-Barshad (ZGB) model,
which was designed in order to describe the catalytic 
reaction CO + O $\rightarrow$ CO$_2$ on a platinum 
surface~\cite{ZGB86}, displays a DP transition. 
In the ZGB model a gas composed of 
CO and O$_2$ molecules with fixed concentrations
$y$ and $1-y$, respectively, is brought into contact 
with a catalytic material. The catalytic surface is 
represented by a square lattice whose sites can either be 
vacant~($\vacancy$), occupied by a CO molecule, 
or occupied by an O atom. 
CO molecules fill any vacant site at rate $y$, whereas
O$_2$ molecules dissociate on the surface into two
O atoms and fill pairs of adjacent vacant sites at rate $1-y$.
Finally, neighboring CO molecules and O atoms recombine
instantaneously to CO$_2$ and desorb from the surface.
On the lattice the three processes correspond to the reaction scheme
\begin{eqnarray}
\vacancy \rightarrow  \text{CO} \hspace{5mm} &&\qquad
\text{at rate $y$ \ ,} \nonumber \\
\vacancy+\vacancy  \rightarrow  \text{O}+\text{O} &&\qquad
\text{at rate $1-y$ \ ,}  \\
\text{O} + \text{CO} \rightarrow \vacancy + \vacancy &&\qquad
\text{at rate $\infty$ \ .} \nonumber
\end{eqnarray}
Therefore, if the whole lattice  is entirely covered
either with CO or O, the system is
trapped in a catalytically inactive state. These `poisoned'
states are the two absorbing configurations of the ZGB model.
As shown in Fig.~\ref{FIGZGB}, the corresponding phase 
diagram displays two absorbing phases.
For $y<y_1 \simeq 0.389$ the system evolves
into the O-poisoned state whereas
for $y>y_2 \simeq 0.525$ it always reaches the CO-poisoned
state. Between these two values the model is
catalytically active. The two transitions into the absorbing
phases are different in character, namely
discontinuous at $y=y_2$ and continuous at $y=y_1$ 
(see Fig.~\ref{FIGZGB}). Motivated by the DP conjecture,
Grinstein et~al.~\cite{GLB89} expected
the latter to belong to the DP universality class.
In order to verify this hypothesis,
extensive numerical simulations were performed.
Initially it was believed that the critical exponents
were different from those of DP~\cite{Meakin90}, while later the
transition at $y=y_1$ was found to belong to DP~\cite{JFD90}.
Very precise estimates of the critical exponents were recently 
obtained in Ref.~\cite{VoigtZiff97}, confirming the DP conjecture.
DP exponents were also obtained in a simplified version of the
ZGB model~\cite{ABW90}. These theoretical models therefore 
suggest that certain catalytic reactions could serve as an 
experimental realization of DP.

%
%
%
\begin{table}
\small
\begin{center}
\begin{tabular}{||c||c|c|c|c||}
exponent
&	  $d=1$~\cite{Jensen99} &
	  $d=2$~\cite{VoigtZiff97} &
	  $d=3$~\cite{Jensen92} &
	  $d\geq 4$ \\ \hline

$\beta$ & $0.276486(8)$ & 0.584(4) & 0.81(1) &  $1$ \\
$\nu_\perp$ & $1.096854(4)$ & 0.734(4) & 0.581(5) & $1/2$ \\
$\nu_\parallel$ & $1.733847(6)$ & 1.295(6) & 1.105(5) &  $1$ 

\end{tabular}
\end{center}
\caption{
\mark{TABEXP}
Numerical estimates for the critical exponents of directed 
percolation in $d$+1 dimensions.
}
\end{table}

In real catalytic reactions, however, only
the discontinuous transition at $y=y_2$ can be observed.
The schematic graph on the right-hand side of Fig.~\ref{FIGZGB} 
shows the reaction rates as functions 
of the CO pressure measured in a
catalytic reaction on a Pt(210) surface~\cite{EMFBCRH89}.
Although the experiment was designed in order to 
investigate the technologically interesting
regime of high activity close to the first-order phase
transition, it clearly indicates that poisoning with oxygen
does not occur. Instead the reactivity increases almost
{\it linearly} with the CO pressure. 
Similar results were obtained for Pt(111) and for other
catalytic materials. Thus, so far there is no
experimental evidence for the DP transition predicted 
by the ZGB model.

%
%

One may speculate why the DP transition 
is obscured or even destroyed under experimental conditions.
A possible reason might be the reaction chain
being much more complicated than in the ZGB model~\cite{EKE90}.
Moreover, the O-poisoned system might not be in a perfect 
absorbing state, i.e., the surface can still adsorb CO 
molecules although it is satured with carbonmonoxide.
Another possibility is thermal (nonreactive) desorption of oxygen
which -- in the DP language -- would correspond to 
spontaneous creation of active sites 
due to an external field~\cite{DickmanPrivate}.
Finally, defects and inhomogeneities of the catalytic 
material could lead to an effective (spatially quenched) disorder.
As shown in the Appendix, this type of disorder is marginal.
It can therefore seriously affect the critical behavior
and even modify the values of the exponents. 

For a long time microscopic details were difficult to study 
experimentally. However, novel techniques
such as scanning tunneling microscopy (STM) led to an
enormous progress in the understanding of catalytic
reactions. They also point at various unexpected
subtleties. For example, recent experiments revealed that
the reactions preferably take place at the perimeter of
oxygen islands~\cite{WVJZE97}. Furthermore, it was observed 
that the adsorbed CO molecules on Pt(111)
may form three different rotational patterns representing the
c(4$\times$2) structure of CO on platinum, leading to
three competing absorbing states~\cite{Hinrichsen97}. 
Moreover, the STM technique allows one to trace individual molecular
reactions and to determine the corresponding reaction rates. 
In addition, the influence of defects such
as terraces on catalytic reactions can be 
quantified experimentally~\cite{ZWTE96}. We may therefore expect a 
considerable progress in the understanding of catalytic 
reactions in near future.

\section{Growing interfaces}
\label{GrowthSection}

Various models for interface growth exhibit a roughening
or depinning transition which can be related to DP.
In this Section we discuss three examples, namely transitions
of depinning interfaces in random media, polynuclear growth
processes, and solid-on-solid growth with evaporation
at the edges of plateaus.

\subsection{Depinning transitions}
\label{DepinningSubsection}

Depinning transitions of driven interfaces
provide a very promising class of experiments 
which could be related to DP~\cite{BarabasiStanley95}.
In these experiments a liquid is pumped through a porous medium.
If the driving force $F$ is sufficiently low the liquid cannot 
move through the medium since the air/liquid interface is
pinned at certain pores. Above a critical threshold,
however, the interface starts moving through
the medium with an average velocity $v$. Close to the transition,
$v$ is expected to scale as
\begin{equation}
v \sim \biggl( \frac{F-F_c}{F_c} \biggr)^\theta \ ,
\end{equation}
where $\theta$ is the velocity exponent. Moreover, in the moving
phase $F>F_c$ the interface roughness averaged over length $\ell$
should obey the usual scaling law for roughening 
interfaces~\cite{FamilyVicsek85}
\begin{equation}
\label{GrowthScaling}
w(\ell,t) \sim \ell^\alpha f(t/\ell^{\tilde{z}}) \,,
\end{equation}
where $\alpha$ is the so-called roughening exponent.
One of the first experiments in 1+1 dimensions was performed by
Buldyrev et~al., who studied the wetting of paper in a basin
filled with suspensions of ink or coffee~\cite{BBCHSV92}.
Measuring the interface width they found the roughness
exponent $\alpha=0.63(4)$. In various other experiments
the values are scattered between $0.6$ and $1.25$. This
is surprising since the Kardar-Parisi-Zhang (KPZ)
class~\cite{KPZ86,HalpinZhang95}, the canonical universality class
for roughening interfaces, predicts the exponent $\alpha=1/2$ 
which is smaller than the experimentally observed values.

It is believed that the large values of $\alpha$ are due 
to inhomogeneities of the porous medium. Due to these
inhomogeneities, the interface does not propagate uniformly
by local fluctuations as in the KPZ equation, it rather
propagates by {\it avalanches}. In the literature two
universality classes for this type of interfacial growth have been
proposed. In case of {\it linear} growth the interface 
should be described by a random field Ising model~\cite{Nattermann92},
leading to the exponents $\alpha=1$, ${\tilde{z}}=4/3$, and $\theta=1/3$
in 1+1 dimensions. In the presence of a KPZ-type nonlinearity, however, 
the roughening process should exhibit a {\it depinning transition} 
which is related to DP~\cite{TangLeschhorn92,MBLS98}. Notice that the
underlying DP mechanism of depinning transitions differs significantly
from an ordinary directed percolation process in a porous
medium subjected to a gravitational 
field (cf. Sect.~\ref{OtherAppSection}). In the latter 
case the spreading agent is restricted to percolate 
along a given direction, i.e., the flow is strictly unidirectional. 
In depinning experiments, however, water may flow both 
according the pumping force {\it and} -- even more
easily -- in opposite direction, as will be explained below.

%
%
\begin{figure}
\epsfxsize=85mm
\centerline{\epsffile{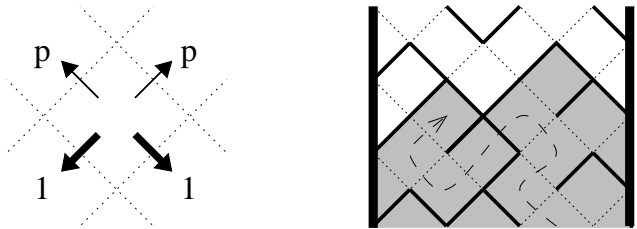}}
\caption{\small
\label{FIGDEPINNINGMODEL}
Simple model exhibiting a depinning transition. The pores are
represented by cells on a diagonal square lattice. The
permeability  across the edges of the cells depends on the
direction of flow: In downwards direction all edges are
permeable whereas in upwards direction they are permeable 
with probability $p$ and impermeable otherwise. 
The right panel of the figure shows a particular configuration
of open (dashed) and closed (solid) edges. 
Pumping in water from below, the interface becomes
pinned along a eirected path of solid lines, 
leading to a finite cluster of wet cells (shaded region).
The dashed arrow represents an open path in order to illustrate
the flow.
}
\end{figure}

A simple model exhibiting a depinning transition is shown
in Fig.~\ref{FIGDEPINNINGMODEL}. In this model the pores are 
represented by cells of a diagonal square lattice. 
The liquid can flow to neighboring cells by crossing 
the edges of the cell. Depending on the direction of the flow
these edges can either be open or closed. For simplicity we assume that 
all edges are permeable in downwards direction, whereas in upwards
direction they can only be crossed with a certain probability $p$.
Thus, by starting with a horizontal row of wet cells at the bottom, 
we obtain a compact cluster of wet cells,
as illustrated in Fig.~\ref{FIGDEPINNINGMODEL}.
The size of the cluster (and therewith the penetration depth of the liquid) 
depends on $p$. If $p$ is large enough, 
the cluster is infinite, corresponding
to a moving interface. If $p$ is sufficiently small, the cluster 
is bound from above, i.e., the interface becomes pinned. 

\begin{figure}
\epsfxsize=85mm
\centerline{\epsffile{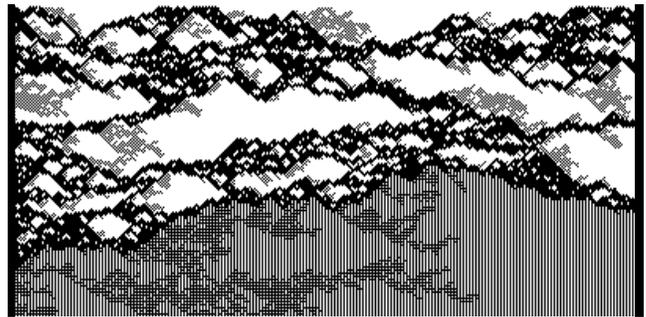}}
\caption{\small
\label{FIGDEPINNING}
Pinned interface and the underlying DP process.
The figure is explained in the text.
}
\end{figure}

The depinning transition is related to DP as follows. 
As can seen in the figure, a pinned interface 
may be interpreted as a {\it directed} path along impermeable 
edges running from one boundary of the system 
to the other. Obviously, the interface becomes pinned only if 
there exists a directed path of impermeable bonds
connecting the boundaries of the system. Hence the depinning 
transition is related to an underlying DP process
running {\it perpendicular} to the direction of growth.
The pinning mechanism is illustrated in Fig.~\ref{FIGDEPINNING},
where a supercritical DP cluster propagates 
from left to right. The cluster's {\em backbone}, 
consisting bonds connecting the two boundaries, 
is indicated by bold dots. The
shaded region denotes the resulting cluster of wet cells.
As can be seen, the interface will be pinned at the lowest
lying branch of the DP backbone. Therefore, the roughening exponent
coincides with the meandering exponent
\begin{equation}
\label{DepinningExponent}
\alpha = \nu_\perp/\nu_\parallel
\end{equation}
of the backbone. Moreover, by analyzing 
the dynamics of the moving interface, 
it can be shown that the dynamic critical 
exponents are given by $\theta=\alpha$ and ${\tilde{z}}=1$.
Thus, depinning transitions in inhomogeneous porous media may serve as 
experimental realizations of the DP universality class.

%
%

Comparing the prediction (\ref{DepinningExponent}) 
with the result $\alpha=0.63(4)$ obtained 
by Buldyrev et~al.~\cite{BBCHSV92} we find an
excellent coincidence, confirming the validity of the
model introduced above. Therefore, this experiment
can be regarded as a first experimental evidence of
DP exponents. However, only {\em one} exponent has been
verified, and it is not fully clear how accurate and 
reproducible these exponents are. Further experimental 
effort in this direction would be desirable.

Similar experiments were carried out in 2+1 dimensions 
with a spongy-like material used by florists, 
as well as fine-grained paper rolls~\cite{BBHKSX92}.
In this case, however, the exponent $\alpha$ is not
related to 2+1-dimensional DP, instead it corresponds to 
the dynamic exponent of percolating directed interfaces
in 2+1 dimensions. In experiments as well as in numerical
simulations a roughness exponent $\alpha=0.50(5)$ was obtained.

In order to perform a depinning experiment 
which can easily be reproduced, 
Dougherty and Carle measured the dynamical avalanche 
distribution of an air/water interface moving through a
porous medium made of glass beads~\cite{DoughertyCarle98}.
Assuming an underlying DP process, the distribution $P(s)$ 
of avalanche sizes $s$ is predicted to behave algebraically. 
In the experiment, however, a stretched exponential behavior
$P(s) \sim s^{-b} e^{-s/L}$ is observed even for small flowing rates.
The estimates for the exponent $b$ are inconclusive; they depend
on the time window of the measurement and vary between $-0.5$
and $0.85$. Even more recently Albert et~al. proposed a
method allowing to identify the universality class by measuring
the propagation velocity of locally tilted parts of the
interface~\cite{ABCD98}. Their results suggest that interfaces
propagating in glass beads are not described by a DP depinning
process, but to be related instead to the random-field Ising
model. From the theoretical point of view this is surprising since
linear growth of the interface is a special case which requires
fine-tuning of certain parameters. Further experimental effort would 
be needed to understand these findings.

\subsection{Polynuclear growth}

A completely different DP mechanism is responsible for
roughening transitions in so-called polynuclear growth 
(PNG) models~\cite{KerteszWolf89,LRWK90,Toom94}. 
A key feature of PNG models is the use of {\em parallel updates}, leading
to a maximal propagation velocity of one monolayer per time step. 
For a high adsorption rate the interface is smooth, propagating 
at maximal velocity $v=1$. Decreasing the adsorption rate below a
certain critical threshold, PNG models exhibit a 
roughening transition to a rough phase with $v<1$. 
In contrast to equilibrium roughening transitions, which only 
exist in $d \geq 2$ dimensions, PNG models have a
roughening transition even in one spatial dimension.

%
%
\begin{figure}
\epsfxsize=85mm
\centerline{\epsffile{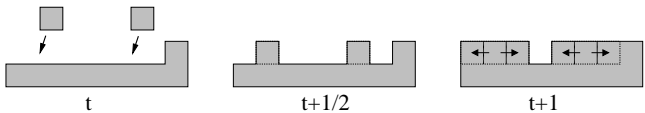}}
\vspace{2mm}
\caption{\small
\label{FIGPNG}
Polynuclear growth model. In the first half time step atoms
are deposited with probability $p$. In the second half time step
islands grow deterministically by one step and coalesce.
}
\end{figure}

Probably the simplest PNG model investigated 
so far is defined by the following dynamic
rules~\cite{KerteszWolf89}.
In the first half time step atoms `nucleate' stochastically
at the surface by
\begin{equation}
\label{nucleation}
h_i(t+1/2) = \left\{
\begin{array}{ll}
h_i(t)+1 \ & \text{with prob. } p \ , \\
h_i(t)   \ & \text{with prob. } 1-p \ . 
\end{array}
\right.
\end{equation}
In the second half time step the islands grow deterministically in
lateral direction by one step. This type of growth may be expressed
by the update rule
\begin{equation}
\label{coalescense}
h_i(t+1) = \max_{j \in <i>} \bigl[ h_i (t+1/2) , h_j (t+1/2) \bigr] \ ,
\end{equation}
where $j$ runs over the nearest neighbors of site $i$. 

The relation to DP can be established as follows. 
Starting from a flat interface $h_i(0)=0$, let us 
interpret sites at maximal height $h_i(t)=t$ as active 
sites of a DP process. The adsorption 
process~(\ref{nucleation}) turns active sites
into the inactive state with probability $1-p$, 
while the process~(\ref{coalescense}) resembles
offspring production. Therefore, if $p$ is large
enough, the interface is smooth and propagates with maximal
velocity $v=1$. This situation corresponds to the active
phase of DP. Therefore, we expect the density of sites at
maximal height to scale as
\begin{equation}
\frac{1}{N} \sum_i \, \delta_{h_i-t} \sim (p-p_c)^\beta,
\end{equation}
where $N$ denotes the system size.
Below a critical threshold, however,
the density of active sites at the maximum height $h_i(t)=t$
vanishes; the growth velocity is smaller than~$1$ 
and the interface evolves into a rough state.
Although this mapping to DP is not exact, numerical simulations
suggest that it still remains valid on a qualitative level. 
More specifically, it turns out that PNG models 
are related to a unidirectionally 
coupled hierarchy of DP processes~\cite{CoupledDP}.

A closely related class of models was introduced in order to describe 
the growth of colonial organisms such as fungi 
and bacteria~\cite{LopezJensen98}, motivated by recent 
experiments with the yeast {\em Pichia membranaefaciens} on
solidified agarose film~\cite{Sams97}. 
By varying the concentration of polluting 
metabolites, different front morphologies
were observed. The model proposed in~\cite{LopezJensen98} 
aims to explain these morphological transitions on a 
qualitative level. A careful analysis of the dynamic rules 
shows that models for fungal growth and PNG models are very 
similar in character. They both employ parallel dynamics 
and exhibit a DP-related roughening transition.

Concerning experimental realizations of PNG models, one 
major problem -- apart from quenched disorder --
is the use of parallel updates. The type of update is crucial;
by using random-sequential updates the transition is lost since
in this case there is no maximum velocity. However, in realistic
experiments atoms do not move synchronously, but the adsorption
events are rather randomly distributed in time. Therefore, random sequential 
updates might be more appropriate to describe such experiments.
It thus remains an open question to what extent PNG processes can
be realized in nature. In fact, it would be interesting to see
if PNG can be generalized to random-sequential dynamics.

\subsection{Growth with evaporation at the edges of plateaus}

DP-related roughening transitions can also be observed in certain
solid-on-solid growth processes with random-sequential 
updates~\cite{AEHM96,AEHM98}. As a key feature of these models, 
atoms may desorb exclusively at the 
{\em edges} of existing layers,
i.e., at sites which have at least one neighbor at a lower height.
By varying the growth rate, such growth processes display a
roughening transition from a non-moving smooth phase to
a moving rough phase. 

A simple solid-on-solid model for this
type of growth is defined by the following dynamic 
rules~\cite{AEHM96}: For each update a site~$i$ 
is chosen at random and an atom is adsorbed
\begin{equation}
\label{adsorption}
h_i \rightarrow h_{i}+1 \; \; \text {with probability } q 
\end{equation}
or desorbed at the edge of a plateau
\begin{equation}
\begin{array}{ll}
\label{desorption}
&h_i \rightarrow \text{min}(h_{i},h_{i+1}) \; \;
\text {with probability } (1-q)/2 \ , \\
&h_i \rightarrow \text{min}(h_{i},h_{i-1}) \; \;
\text {with probability } (1-q)/2 \ .
\end{array}
\end{equation}
Moreover, the growth process is assumed to be {\it restricted}, 
i.e., updates are only carried out if the resulting configuration obeys
the constraint
\begin{equation}
\label{RSOS}
|h_{i}-h_{i\pm 1}| \leq 1 \ .
\end{equation} 
%

%
%
\begin{figure}
\epsfxsize=85mm
\centerline{\epsffile{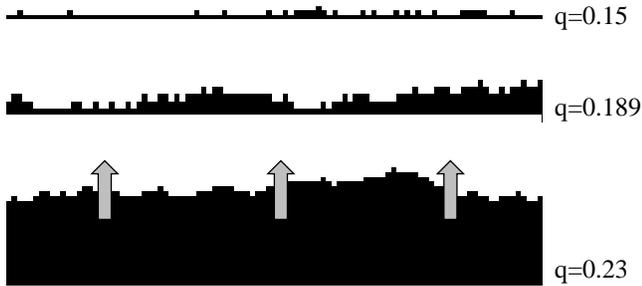}}
\vspace{1mm}
\caption{\small
\label{FIGMONOMER}
Restricted solid-on-solid growth model exhibiting a roughening
transition from a non-moving smooth to a moving rough phase.
Monomers are randomly deposited whereas desorption takes place
only at the edges of plateaus.
}
\end{figure}

The qualitative behavior of this model 
is illustrated in Fig.~\ref{FIGMONOMER}.
For small~$q$ the desorption processes (\ref{desorption}) dominate.
If all heights are initially set to the same value, this level will
remain the bottom layer of the interface. Small islands will grow on 
top of the bottom layer but will be quickly eliminated
by desorption at the island edges.  Thus, the interface is effectively
anchored to its bottom layer and a smooth phase is maintained.
The growth velocity $v$ is therefore zero in the thermodynamic limit.
As $q$~is increased, more islands on top of the bottom layer
are produced until above~$q_c \simeq 0.189$, the critical value of~$q$, they
merge forming new layers at a finite rate, giving rise
to a finite growth velocity. In an infinite system the growth
velocity scales near the transition as
\begin{equation}
v \sim \xi_\parallel^{-1} \sim (p-p_c)^{\nu_\parallel} \,.
\end{equation}
Since the propagation velocity fluctuates locally,
the interface evolves into a rough state according to the predictions
of the KPZ equation. The same type of critical behavior is observed in
similar models without the restriction~(\ref{RSOS}).

At the transition the dynamics of the model is related to DP as follows.
Starting with a flat interface at zero height, let us consider all sites
with $h_i=0$ as particles $A$ of a DP process. Growth according to 
Eq.~(\ref{adsorption}) corresponds to spontaneous annihilation 
$A\rightarrow\vacancy$. Conversely, desorption may be considered as
a particle creation process. However, since atoms may only desorb
at the edges of plateaus, particle creation requires a neighboring
active site, corresponding to offspring production $A \rightarrow 2A$.
These rules resemble (although not exactly) the dynamics of a DP
process. In contrast to PNG models, the DP process takes place at
the bottom layer of the interface. Moreover, the roughening 
transition does not depend on the use of either parallel or 
random-sequential updates. However, if composite particles 
instead of monomers are deposited on the surface, 
the universality class may change due to additional 
symmetries~\cite{HinrichsenOdor99a}.

%
%
\begin{figure}
\epsfxsize=65mm
\centerline{\epsffile{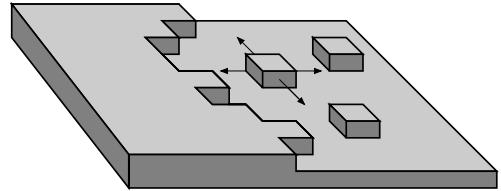}}
\vspace{2mm}
\caption{\small
\label{FIGPERSPECTIVE}
Possible experimental realization of the growth process defined
in Eqs.~(\ref{adsorption}) and (\ref{desorption}) by lateral growth
of a monolayer on a substrate.
}
\end{figure}

With respect to experimental realizations of the dynamic rules 
(\ref{adsorption})-(\ref{desorption}) we note that atoms 
are not allowed to diffuse on the surface. 
This assumption is rather unnatural since in most 
experiments the rate for surface diffusion is much higher
than the rate for desorption back into the gas phase. 
Therefore, it will be difficult to realize this type of 
homoepitaxial growth experimentally. However, in a different
setup, the above model could well be relevant~\cite{OferBiham}.
As illustrated in Fig.~\ref{FIGPERSPECTIVE}, a laterally growing 
monolayer could resemble the dynamic rules~(\ref{adsorption})
and~(\ref{desorption}) by identifying the edge of the monolayer
with the interface of the growth model. In this case 
'surface diffusion', i.e. diffusion of atoms along 
the edge of the monolayer, is highly suppressed. Moreover,
in single-step systems (such as fcc(100) surfaces) 
it would also be possible to implement the restriction~(\ref{RSOS}).

\section{Flowing granular matter}
\label{SandSection}

\begin{figure}
\epsfxsize=45mm
\centerline{\epsffile{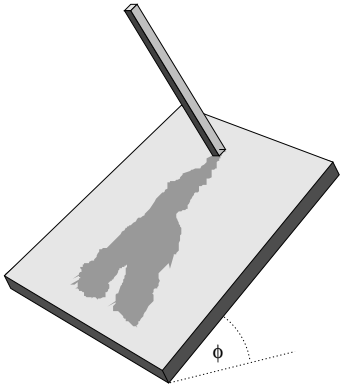}}
\vspace{2mm}
\caption{
Simplified drawing of the Douady-Daerr experiment. 
At a given angle $\phi$ a layer of a certain thickness 
settles. Perturbing the layer locally with a stick leads to
an avalanche of flowing sand.
\label{FigDouadyDaerr}
}
\end{figure}

It has been shown recently that simple systems of
flowing sand on an inclined plane, e.g. the experiments
performed by Douady and Daerr~\cite{DouadyDaerr98,DaerrDouady99},
could serve as experimental realizations of DP~\cite{HJRD99}. 
In the Douady-Daerr experiment glass beads with a diameter 
of $250$-$425 \ \mu$m are poured uniformly 
at the top of an inclined plane covered by a rough velvet cloth 
(see Fig.~\ref{FigDouadyDaerr}). As the beads flow down, 
a thin layer settles and remains immobile. Increasing
the angle of inclination $\phi$ by $\Delta\phi$
the layer becomes dynamically unstable, i.e., 
by locally perturbing the system at the top of the plane
an avalanche of flowing granular matter will be generated. 
In the experiment these avalanches have the shape of a fairly 
regular triangle with an opening angle~$\theta$.
As the increment $\Delta \varphi$ decreases, the value of $\theta$ 
decreases, vanishing as 
\begin{equation}
\tan\theta \sim (\Delta \varphi)^x \,
\label{eq:powerlaw}
\end{equation}
with a certain critical exponent $x$. 
The experimental results suggest the value $x=1$~\cite{DaerrDouady99}.

%
%
%

In order to explain the experimentally observed triangular form of the
avalanches, Bouchaud {\it et al.} proposed a mean-field theory based on
deterministic equations taking the actual local thickness of the 
flowing avalanche into account~\cite{BouchaudCates98}. 
This theory predicts the exponent $x=1/2$. 
Another explanation assumes that flowing sand may be 
associated with a nearest-neighbor spreading process~\cite{HJRD99}. 
Considering the avalanche as a cluster of active sites while 
identifying the vertical coordinate of the plane with time 
and the increment of inclination $\Delta \varphi$ 
with $p - p_c$ the opening angle is expected to scale as
\begin{equation}
\tan \theta \sim \xi_\perp / \xi_\parallel \sim 
(\Delta \varphi )^{\nu_\parallel - \nu_\perp } \,,
\label{eq:Dnu}
\end{equation}
where $\nu_\parallel$ and $\nu_\perp$ are the scaling
exponents of the spreading process under consideration.

To support this scaling argument, a simple lattice model was
introduced which mimics the physics of flowing sand~\cite{HJRD99}.
The model exhibits a transition from an inactive to an 
active phase with avalanches whose compact shapes reproduce 
the experimental observations. On laboratory scales the model 
predicts the exponent $x=1$, corresponding to the universality 
class of {\em compact directed percolation} (CDP) which
is characterized by the exponents~\cite{DickmanTretyakov95}
\begin{equation}
\nu_\parallel=2\,, \qquad \nu_\perp=1 \,, \qquad \beta=0\,.
\end{equation}
The CDP behavior, however, is only an initial transient and
crosses over to DP after a very long time. Thus
the Douady-Daerr experiment -- performed on sufficiently
large scales -- may serve as a physical realization of DP. 
Irregularities of the layers thickness may affect the
spreading properties of avalanches. However, such
inhomogeneities can be considered as a kind of spatio-temporally 
quenched disorder which is irrelevant on large scales
(see Appendix). Thus, in contrast to the previous examples,
the problem of quenched disorder does not play a major 
role in this type of experiments.

%
%
%
\begin{figure}
\epsfxsize=80mm
\centerline{\epsffile{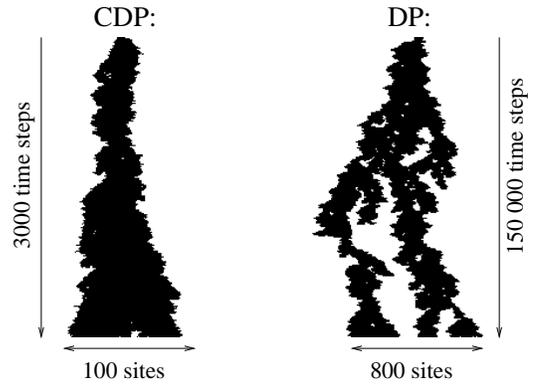}}
\vspace{2mm}
\caption{
Typical clusters generated at 
criticality on small and large scales,
illustrating the crossover from CDP to DP.
\label{FigDemoCrossover}
}
\end{figure}

The crossover from CDP to DP is very slow and presently not 
accessible in the experiments. To illustrate the crossover,
two avalanches are plotted on different scales
in Fig.~\ref{FigDemoCrossover}. The left one 
represents a typical avalanche within the first
few thousand time steps. As can be seen, the cluster appears 
to be compact. However, as shown in the right panel 
of Fig.~\ref{FigDemoCrossover}, the cluster breaks up 
into several branches after a very long time. As a precondition 
for DP behavior, initially compact avalanches should thus be 
able to break up into several branches. Only then is it 
worthwhile to optimize the experimental setup and to measure 
the critical exponents quantitatively.

\begin{figure}
\epsfxsize=65mm
\centerline{\epsffile{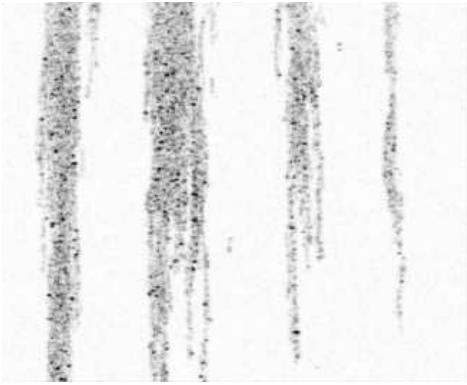}}
\vspace{1mm}
\caption{\small
\label{FIGSPLIT}
Avalanches splitting up into several branches,
observed in recent experiments with high angle
of inclination (reprinted with kind permission from
A. Daerr).
}
\end{figure}

More recent experimental studies~\cite{Daerr99} confirm that for
high angles of inclination critical avalanches do split
up into several branches (see Fig.~\ref{FIGSPLIT}). 
Yet here the avalanches have no well defined front,
the propagation velocity of separate branches rather depends 
on their thickness. It is therefore no longer possible to interpret 
the vertical axis as a time coordinate. Moreover, it is not yet 
known how the spreading process depends on correlations in the 
initial state. As shown in Ref.~\cite{HinrichsenOdor98}, such long-range 
correlations may change the values of certain dynamic critical 
exponents. However, recent studies of a single rolling grain on 
an inclined rough plane~\cite{DBW96} support that there are presumably no 
long-range correlations due to a `memory' of rolling grains.  
By means of molecular dynamics simulations it was shown that 
the motion of a rolling grain consists of many small bounces 
on each grain of the supporting layer. Therefore, the rolling
grain quickly dissipates almost all of the energy gain from the 
previous step and thus forgets its history very fast. For this reason
it seems to be unlikely that quenched disorder of the prepared 
layer involves long-range correlations. Therefore, flowing granular 
matter seems to  be a promising candidate for an experimental 
realization of DP.

%
%
%

\section{Other Applications}
\label{OtherAppSection}

This section discusses other applications which are less 
promising to serve as experimental realizations of DP,
although they are frequently quoted in the literature. 


{\bf Porous media:}
One example is percolating water in a porous medium subjected to an
external driving force. The medium could be a porous rock in
a gravitational field where neighboring pores are connected 
by channels with a certain probability. Depending on this probability,
the penetration depth is either finite or the water may 
"percolate" over infinitely long distances through the medium.
Due to the external driving force, the flow in the medium is assumed to be
strictly unidirectional, i.e., the water can only flow downwards
(in contrast to the depinning models of Sect.~\ref{DepinningSubsection}).
Although this application is quite natural, it is extremely difficult
to realize experimentally. For example, by studying natural sandstone
a broad distribution of pore sizes was observed\cite{HRV97}. 
Although these experiments are concerned with isotropic percolation,
similar difficulties are expected in the directed case.


{\bf Epidemics:}
Another frequently quoted applications of DP is the spreading 
of epidemics without immunization~\cite{Harris74,Mollison77,Liggett85}. 
Here infection and recovery resemble 
the reaction-diffusion scheme~(\ref{DPReactionDiffusion}).
If the rate of infection is very low, the infectious disease
will disappear after some time. If infections occur more 
frequently, the disease may spread and survive for a very 
long time. However, spreading processes in nature are usually not
homogeneous enough to reproduce the critical behavior of DP.
Moreover, in many realistic spreading processes short-range
interactions are no longer appropriate. This situation emerges, 
for example, when an infectious disease is transported by insects. 
The motion of the insects is typically not a random walk, 
rather occasional flights over long distances may occur
before the next infection takes place. 
On the level of theoretical models such long-range interactions
may be described by L\'evy flights~\cite{BouchaudGeorges90}, leading to 
continuously varying critical exponents~\cite{JOW99}.


{\bf Forest fires:}
A closely related problem is the spreading of forest 
fires~\cite{Albano94a}. Tephany {\it et al.} studied the
propagation of flame fronts on a random lattice
both under quiescent conditions and in a wind tunnel~\cite{TND97}.
The experimental estimates of the critical exponents at the 
spreading transition are in rough agreement with the 
predictions of isotropic and directed percolation, respectively.
However, the accuracy of these costly experiments remains limited.


{\bf Calcium dynamics:}
DP transitions may also occur in certain kinetic models for the
dynamics of Calcium ions in living cells. Ca$^{2+}$ ions play an
important physiological role as second messenger for various purposes
ranging from hormonal release to the activation of egg cells by
fertilization~\cite{Berridge93,KRS95}. The cell uses nonlinear 
propagation of increasing intracellular Ca$^{2+}$ concentration,
a so-called calcium wave, as a tool to transmit 
signals over distances which are much longer than the diffusion length.
For example, propagating Ca$^{2+}$ waves can be observed in the
immature {\it Xenopus laevis oocyte}~\cite{LechleiterClapham92}.
So far theoretical work focused mainly on deterministic
reaction-diffusion equations in the continuum which explain various
phenomena such as solitary and spiral waves~\cite{ReviewWaves}.
This mean-field type approach, however, 
ignores the influence of fluctuations.
Yet, near the transition between survival and extinction of
Ca$^{2+}$, activity fluctuations may play an important role.
Recently improved models have been introduced which 
take also the stochastic nature of Calcium release into
account~\cite{KeizerSmith98,BFLT}. As expected, 
the transition in one of these models
belongs to the DP universality class~\cite{BFLT}.
However, from the experimental point of view it seems to be
impossible to confirm or disprove this conjecture. On the one hand,
the size of a living cell is only a few order of magnitude larger
than the diffusion length, leading to strong finite-size effects
in the experiment. On the other hand, inhomogeneities as well as internal
structures of the cell give rise to a completely unpredictable form
of quenched noise which may be {\it correlated} in space and time.
Therefore, it seems to be impossible to identify the universality
class of the transition in actual experiments. It would be rather
an achievement to find clear evidence for the very existence of a phase
transition between survival and extinction of propagating calcium
waves. For quantitative experiments, it would be interesting to
reproduce the dynamics of Calcium in a well-defined environment,
e.g., on an artificial membrane~\cite{DickmanPrivate}.


{\bf Directed polymers:}
DP is also related to the problem of directed polymers~\cite{HH85}.
In contrast to DP, which is defined as a {\em local} process, the directed
polymer problem selects directed paths in a random medium
by {\em global} optimization. Under certain conditions, namely
a bimodal distribution of random numbers, both problems were shown
to be closely related~\cite{PerlsmanHavlin99}. More specifically,
the roughness exponent of the optimal path in a directed polymer
problem is predicted to cross over from the KPZ value 2/3 to 
the DP value $\nu_\parallel/\nu_\perp \simeq 0.63$ at
the transition point. Directed polymers were used 
to describe the propagation of cracks~\cite{KHW92}.
However, it is rather unlikely that crack experiments can reproduce
the tiny crossover from KPZ to DP.


{\bf Turbulence:}
Finally, DP has also been considered as a toy model for 
turbulence. According to Ref.~\cite{Pomeau86}, the front 
between turbulent and laminar flow should exhibit the critical
behavior of DP. For example, the velocity of the front should 
scale algebraically with a combination of DP exponents. 
However, these predictions are based rather on heuristic arguments
than on rigorous results. In fact, in many respects turbulent
phenomena show a much richer behavior than DP.

\section{Conclusions}

Directed percolation has kept theoretical physicists 
fascinated since more than forty years. Several 
reasons make directed percolation so appealing.
First of all, DP is a very simple model in terms of its dynamic
rules. Nevertheless, the DP phase transition turns out to
be highly nontrivial. In fact, DP belongs 
to the very few critical phenomena which have not 
yet been solved exactly in one spatial dimension. Therefore, 
the critical exponents are not yet known analytically. 
High-precision estimates indicate that they might be given
rather by irrational than by simple fractional values.

Moreover, DP is extremely robust. It stands for a whole
universality class of phase transitions from a fluctuating phase
into an absorbing state. In fact, a large variety of models 
displays phase transitions belonging to the DP universality 
class. Thus, on the theoretical level, DP plays the role of a
standard universality class similar to the Ising model 
in equilibrium statistical mechanics.

In spite of its simplicity, no experiment is known  
confirming the values of the critical exponents quantitatively.
An exception may be the wetting experiment performed by Buldrey et al. 
where the value of the roughness exponent $\alpha$ coincides with
$\nu_\perp/\nu_\parallel$ within less than 10\%. 
However, since the results of similar experiments are scattered
over a wide range,  further experimental effort would be 
needed in order to confirm the existence of DP in this type of
systems.

Apart from difficulties to realize a non-fluctuating absorbing
state, a fundamental problem of DP experiments is the emergence
of quenched disorder due to certain inhomogeneities of the system. 
Depending on the type of disorder, even weak inhomogeneities 
might obscure or even destroy the DP transition.
Therefore, the most promising experiments are those where
quenched disorder is irrelevant on large scales. This is the case,
for example, in wetting experiments (Sect.~\ref{DepinningSubsection})
and systems of flowing granular matter (Sect.~\ref{SandSection}).

In spite of all these difficulties, many physicists believe that
DP should have a counterpart in reality, mostly because 
of its simplicity and robustness. Therefore, Grassbergers message
remains valid: The experimental realization of DP is an outstanding
problem of top priority.

\vspace{4mm}
\noindent
Acknowledgements:
I would like to thank R. Dickman for carefully reading the
manuscript. I would also like to thank A. Daerr and O. Biham
for fruitful suggestions and discussions.

\appendix
\section{Quenched disorder}
\renewcommand{\thesubsection}{\Alph{subsection}}

On a coarse-grained scale the temporal evolution of a DP process
without quenched disorder is described by the
Langevin equation~\cite{DPConjecture}
\begin{equation}
\partial_t \rho = a\rho - \lambda\rho^2 + D \nabla^2\rho + \sqrt{\rho}\zeta\,,
\end{equation}
where $\zeta(\xvec,t)$ denotes an uncorrelated Gaussian noise:
\begin{equation}
\langle \zeta(\xvec,t)\zeta(\xvec',t')\rangle =
\Gamma \delta^d(\xvec-\xvec')\delta(t-t') \,.
\end{equation}
The noise $\zeta(\xvec,t)$
represents the {\em intrinsic} fluctuations of the DP process due to the 
stochastic nature of the dynamic rules. The parameter $a$ controls
the rate for offspring production and can be thought of as
a being measure of the percolation probability $p-p_c$. 

Quenched disorder may be introduced by 
random variations of the parameter $a$, i.e.,
by adding another noise field $\eta$:
\begin{equation}
\label{Replacement}
a \rightarrow a + \eta \,.
\end{equation}
Thus, the resulting Langevin equation reads
\begin{equation}
\label{LangevinEquation}
\partial_t \rho = a\rho - \lambda\rho^2 + D \nabla^2\rho + \sqrt{\rho}\zeta
+ \rho\eta \,.
\end{equation}
The noise $\eta$ is {\em quenched} in the sense
that quantities like the particle density 
are averaged over many independent realizations 
of the intrinsic noise $\zeta$ while 
the disorder field $\eta$ is kept fixed.
In the following we distinguish three different types of
quenched disorder:
\begin{enumerate}

	\item[A.] Spatially quenched disorder $\eta_s(\xvec)$.

	\item[B.] Temporally quenched disorder $\eta_t(t)$.

	\item[C.] Spatio-temporally quenched disorder $\eta_{st}(\xvec,t)$.

\end{enumerate}
These variants of quenched disorder
differ in how far they affect the critical 
behavior of DP. In the following we review some
of the main results.

\subsection{Spatially quenched disorder}

For spatially quenched disorder, the noise field $\eta$ is defined
through the correlations
\begin{equation}
\overline{\eta_s(\xvec)\eta_s(\xvec')}=\gamma\,\delta^d(\xvec-\xvec') \,,
\end{equation}
where the bar denotes the average over independent realizations 
of the disorder field (in contrast to averages $\langle\ldots\rangle$
over the intrinsic noise $\zeta$). 
The parameter $\gamma$ is an amplitude which controls the intensity
of disorder. In order to find out whether this type of noise affects the
critical behavior of DP, let us consider the properties of 
the Langevin equation under the scaling transformation
\begin{equation}
\label{Rescaling}
\xvec \rightarrow \Lambda \xvec \,, \qquad
t \rightarrow \Lambda^z t \,, \qquad
\rho \rightarrow \Lambda^{-\chi} \rho\,,
\end{equation}
where $\Lambda$ is a dilatation factor while
$z=\nu_\parallel/\nu_\perp$ and $\chi = \beta/\nu_\perp$ 
are the critical exponents of DP. In absence of quenched disorder, 
the Langevin equation turns out to be invariant under rescaling if
\begin{equation}
\label{MeanFieldExponents}
z=2\,, \qquad \chi=2\,, \qquad
a=0\,, \qquad d=d_c \nonumber \,,
\end{equation}
where $d_c=4$ is the upper critical dimension of DP. 
These values are consistent with the 
DP mean field exponents $\beta=1$, 
$\nu_\perp=1/2$, and $\nu_\parallel=1$,
which are valid in $d \geq 4$ dimensions. Checking
the scaling behavior of the additional term $\rho\eta_s$ 
in Eq.~(\ref{LangevinEquation}) at the critical dimension,
we observe that it scales as
\begin{equation}
\rho\eta_s \rightarrow \Lambda^{-d_c/2-\chi} \rho\eta_s\,,
\end{equation}
i.e., spatially quenched disorder is a {\em marginal}
perturbation. Therefore, it may seriously affect the
critical behavior at the transition.

The same result is obtained by considering the 
field-theoretic action. Without quenched noise, DP
is described by the action of Reggeon field theory~\cite{Reggeon}
\begin{equation}
\label{DPAction}
S_0 = \int d^dx \int dt \, \bar{\psi} 
\Bigl[\partial_t -a - D\nabla^2 + g(\psi-\bar{\psi})\Bigr] \psi
\end{equation}
where $\psi(\xvec,t)$ represents the local particle density while
$\bar{\psi}(\xvec,t)$ denotes the Martin-Siggia-Rosen response field.
As shown by Janssen~\cite{Janssen97}, spatially quenched noise
can be taken into account by adding the term
\begin{equation}
S=S_0 + \gamma \int d^dx \left[ \int dt\,\bar{\psi}\psi \right]^2 \,.
\end{equation}
By simple power counting we can prove that this additional term is 
indeed a marginal perturbation. Janssen showed by a field-theoretic
analysis that the stable fixed point is shifted to an unphysical
region, leading to runaway solutions of the flow equations 
in the physical region of interest. Therefore, spatially quenched
disorder is expected to crucially disturb the critical behavior
of DP. The findings are in agreement with earlier numerical results
by Moreira and Dickman~\cite{MoreiraDickman96} who reported
non-universal logarithmic behavior instead of power laws. 
Later Cafiero {\it et al.}~\cite{CGM98} showed that DP with spatially
quenched randomness can be mapped onto a non-Markovian spreading
process with memory, in agreement with previous results.

From a more physical point of view, spatially 
quenched disorder in 1+1 dimensional systems
was studied by Webman {\it et al.}~\cite{WACH98}. It turns out that
even very weak randomness drastically modifies the phase diagram.
Instead of a single critical point one obtains a whole phase of
very slow glassy-like dynamics. The glassy phase is characterized
by non-universal exponents which depend on the percolation probability
and the disorder amplitude. For example, in a supercritical 1+1 
dimensional DP process without quenched disorder the boundaries
of a cluster propagate at constant velocity $v$. However, in the
glassy phase $v$ decays {\em algebraically} with time. The corresponding
exponent turns out to vary continuously with 
the mean percolation probability. The power-law
behavior is due to `blockages' at certain sites where the
local percolation probability is small. Similarly,
in the subcritical edge of the glassy phase, the spreading
agent becomes localized at sites with high percolation probability.
In $d>1$, however, numerical simulations indicate that a glassy
phase does not exist.

\subsection{Temporally quenched disorder}

Temporally quenched disorder is defined by the correlations
\begin{equation}
\overline{\eta_t(t)\eta_t(t')}=\gamma\,\delta(t-t') \,.
\end{equation}
In this case the additional term scales as a {\em relevant} perturbation
$\rho\eta_t \rightarrow \Lambda^{-z/2-\chi} \rho\eta_t$. 
Therefore, we expect the critical behavior and the associated
critical exponents to change entirely. In the field-theoretic 
formulation this corresponds to adding a term of the form
\begin{equation}
S=S_0 + \gamma \int dt \left[ \int d^dx\,\bar{\psi}\psi \right]^2 
\end{equation}
The influence of spatio-temporally quenched disorder was investigated
in detail by I. Jensen~\cite{Jensen96a}. Employing series expansion
techniques he demonstrated that the three exponents
$\beta, \nu_\perp, \nu_\parallel$ vary continuously with the
disorder strength. Thus the transition no longer belongs to
the DP universality class. A field-theoretic explanation of
these findings is still missing.

\subsection{Spatio-temporally quenched disorder}

For spatio-temporally quenched disorder, 
the noise field $\eta$ is uncorrelated in both space and time:
\begin{equation}
\overline{\eta_{st}(\xvec,t)\eta_{st}(\xvec',t')}=
\gamma\,\delta^d(\xvec-\xvec')\delta(t-t')\,.
\end{equation}
In Reggeon field theory, this would correspond to the addition of the term
\begin{equation}
S=S_0 +\gamma \int d^dx dt \left[\bar{\psi}\psi \right]^2
\end{equation}
which is an {\em irrelevant} perturbation. 
Spatio-temporally quenched disorder is expected in systems where
each time step is associated with a new set of spatial degrees
of freedom. Examples include water in porous media subjected
to a gravitational field as well as flowing sand on an
inclined plane. In these systems the critical behavior of DP
should remain valid on large scales.


%

\end{document}